\documentclass[pdftex,preprint,amsmath,amsfonts,amssymb,a4paper,eqsecnum,
superscriptaddress,longbibliography]{revtex4-1}

\usepackage{hyperref}
\hypersetup{backref,  
colorlinks=true,
citecolor=cyan,
urlcolor=black
}
\expandafter\let\csname equation*\endcsname\relax 
 \expandafter\let\csname endequation*\endcsname\relax 
\usepackage[usenames,dvipsnames]{color}
\usepackage{color,graphicx,amsthm,amsmath,amsfonts}
\usepackage[utf8]{inputenc}

\usepackage[normalem]{ulem}

\usepackage[T1]{fontenc}

\begin{document}

\title{Nonextensive thermodynamic functions in
the Schrödinger-Gibbs ensemble}
\author{J. L. Alonso}
\affiliation{Departamento de F{\'{\i}}sica Te\'orica, Universidad de Zaragoza, 
Pedro Cerbuna 12, E-50009 Zaragoza, Spain}
\affiliation{Instituto de Biocomputaci\'on y F{\'{\i}}sica de Sistemas 
Complejos (BIFI), Universidad de Zaragoza, Mariano Esquillor s/n, Edificio 
I+D, E-50018 Zaragoza, Spain}
\affiliation{Unidad Asociada IQFR-BIFI}
\author{A. Castro}
\affiliation{BIFI-Fundaci\'on ARAID, Universidad de Zaragoza,  Edificio I+D-Campus
  R\'{\i}o Ebro, Mariano Esquillor s/n, 50018 Zaragoza (SPAIN)} 
\author{J. Clemente-Gallardo}
\affiliation{Departamento de F{\'{\i}}sica Te\'orica, Universidad de Zaragoza, 
Pedro Cerbuna 12, E-50009 Zaragoza, Spain}
\affiliation{Instituto de Biocomputaci\'on y F{\'{\i}}sica de Sistemas 
Complejos  (BIFI), Universidad de Zaragoza, Mariano Esquillor s/n, Edificio 
I+D, E-50018 Zaragoza, Spain}
\affiliation{Unidad Asociada IQFR-BIFI}
\author{J. C. Cuch\'{\i}}
\affiliation{Departament d'Enginyeria Agroforestal, ETSEA-Universitat de 
Lleida, Av. Alcalde Rovira Roure 191, 25198 Lleida, Spain}
\author{P. Echenique}
\affiliation{Instituto de Qu\'{\i}mica F\'{\i}sica Rocasolano, CSIC, Serrano 
119, E-28006 Madrid, Spain}
\affiliation{Departamento de F{\'{\i}}sica Te\'orica, Universidad de Zaragoza, 
Pedro Cerbuna 12, E-50009 Zaragoza, Spain}
\affiliation{Instituto de Biocomputaci\'on y F{\'{\i}}sica de Sistemas 
Complejos (BIFI), Universidad de Zaragoza, Mariano Esquillor s/n, Edificio
I+D, E-50018 Zaragoza, Spain}
\affiliation{Unidad Asociada IQFR-BIFI}
\author{J. G. Esteve}
\affiliation{Departamento de F{\'{\i}}sica Te\'orica, Universidad de Zaragoza, 
Pedro Cerbuna 12, E-50009 Zaragoza, Spain}
\affiliation{Instituto de Biocomputaci\'on y F{\'{\i}}sica de Sistemas 
Complejos (BIFI), Universidad de Zaragoza, Mariano Esquillor s/n, Edificio 
I+D, E-50018 Zaragoza, Spain} 
\author{F. Falceto}
\affiliation{Departamento de F{\'{\i}}sica Te\'orica, Universidad de Zaragoza, 
Pedro Cerbuna 12, E-50009 Zaragoza, Spain}
\affiliation{Instituto de Biocomputaci\'on y F{\'{\i}}sica de Sistemas 
Complejos (BIFI), Universidad de Zaragoza, Mariano Esquillor s/n, Edificio 
I+D, E-50018 Zaragoza, Spain}

\begin{abstract}
Schrodinger suggested that thermodynamical functions cannot be based
on the gratuitous allegation that  quantum-mechanical levels
(typically the orthogonal eigenstates of the Hamiltonian operator) are
the only  allowed states for a quantum system [E. Schrodinger, 
Statistical Thermodynamics (Courier Dover, Mineola, 1967)].
Different authors have interpreted this statement by
introducing density distributions on the space of quantum pure states with
weights obtained as functions of the expectation value of the Hamiltonian of
the system.

In this work we focus on one of the best known of
these distributions, and we prove that, when considered in composite
quantum systems, it defines partition functions that do not factorize as products of
partition functions of the noninteracting subsystems, even in the thermodynamical regime. This
implies that it is not possible to define extensive thermodynamical magnitudes
such as the free energy, the internal energy or the thermodynamic entropy by
using these models.   Therefore, we conclude that this distribution  
inspired by Schrödinger's idea can not be used to construct an appropriate
quantum equilibrium thermodynamics.

\end{abstract}

\maketitle

\section{Introduction}

In a note to the second edition of his book on statistical thermodynamics
\cite{schrodinger1989statistical}, Schrödinger suggests that thermodynamical
functions cannot be based on the gratuitous allegation that quantum-mechanical
levels (typically the orthogonal eigenstates of the Hamiltonian operator) are
the only allowed states in statistical thermodynamics. In Khinchin's classical
book on the mathematical foundations of quantum statistics (sec.~2 of chap~III
of \cite{khinchin1998mathematical}) an approach to microcanonical averages is
proposed in line with Schrödinger's suggestion. Landau and Lifshitz
\cite{landau1959statistical} also consider unrealistic the possibility of
preparing isolated macroscopic physical systems in a precise energy eigenstate,
since energy levels are too close to each other to select just one.

From these points of view, it seems then natural to consider the description of
quantum statistical systems in terms of  probability densities defined on the space of
physical states. This description offers the possibility of considering
the weight of a given state  in an
analogous manner to what it is done when describing classical statistical
systems. The approach also offers many advantages in the description of hybrid
quantum-classical systems, as it can be seen in \cite{Alonso2011,Alonso2012a}.
Nonetheless, it is important to keep in mind that these distributions are
ambiguous from the physical point of view. Indeed, it is well known since von
Neumann \cite{von1955mathematical} that the physical properties of a quantum
system are encoded univocally in its density matrix $\hat \rho$, but there are
many different equivalent expressions for $\hat{\rho}$ as a sum of pure states
(projectors). Therefore, we may define entirely different distributions on the
space of states, which nevertheless are equivalent because they lead to the
same density matrix.

In Refs. \cite{Brody1998b,Bender2005a,Brody2007b} Brody et al., propose their
quantum microcanonical postulate which asserts that every quantum state
possessing the same energy expectation value must be realized
 with the same probability.
Brody and coworkers also introduce an alternative quantum canonical
distribution and studied some of its properties. The distribution is defined as a
density on the space of pure quantum states that assigns to each state the
Boltzmann weight associated with the expectation value of the Hamiltonian. These
ideas have been adopted with different degrees of intensity by several authors
in the last years using even different distributions (see
Refs.~\cite{jona2000invariant,jona2006statistics,Naudts2006,Goldstein2006a,Fresch2010,Fresch2009a,Fine2009,Fine2012,Ji2011,Muller2011,Campisi2013a,CamposVenuti2013,Gutkin2013} for some
examples). In this work we focus on the particular state-space
distribution introduced in Refs.
\cite{Brody1998b,Brody2007b,Bender2005a,jona2000invariant,jona2006statistics} 
and, following Jona-Lasinio and Presilla,  we  refer to it as
the Schrödinger-Gibbs (SG) distribution. 
A crucial result of this
construction is that, as suggested by Fig.~2 of Ref.~\cite{Brody1998b}, the
third law of thermodynamics may not be satisfied. Therefore it is desirable to
know why this happens and whether some other essential properties of the
equilibrium thermodynamics, such as the definition of consistent extensive
magnitudes, are maintained when using the  SG distribution.

In this work we prove that, when considered on composite
noninteracting   quantum systems, this
distribution defines partition functions that do not factorize as products of
partition functions of the subsystems, even in the thermodynamical regime. This
implies that it is not possible to define extensive thermodynamical magnitudes
such as the free energy, the internal energy or the thermodynamic entropy by
using these models. Therefore, although Schrödinger's suggestion might seem
reasonable a priori and it could have interesting dynamical features (as can be 
learned from some of the references above), it cannot be used to construct an 
appropriate quantum thermodynamics; at least not if the suggestion is 
materialized in the form of the SG distribution used as an equilibrium
distribution.

The structure of the paper is as follows. In Section
\ref{sec:two-canon-ensembl} we review the traditional canonical distribution as
well as the SG distribution, and we discuss the different alternative
forms to present them. Section \ref{sec:quant-prop-boltzm} studies the quantum
properties associated with  the density matrix that encodes the properties of the
SG distribution, in particular that it commutes with
the Hamiltonian. Section \ref{sec:therm-prop-boltzm} and Section
\ref{sec:orig-non-extens} contain the main result of the paper: We prove that
the partition function of the new quantum ensemble does not factorize in the
case of  composite noninteracting systems. In Section \ref{sec:additivity-entropy} we present a
simple model to exemplify our analysis, and discuss it from several points of
view. Namely, we evaluate the partition function, the possible definitions of
thermodynamic extensive magnitudes, and an alternative description in terms of
density matrices. In both approaches we identify the lack of factorizability
and  one of  its causes: basically, that the density matrix associated with the SG
distribution represents an entangled state. In Section \ref{sec:conclusions}, we
briefly summarize the content of the paper and consider the potential impact of
our analysis on the study of the equilibrium and non-equilibrium statistics
of hybrid quantum-classical systems.

\section{On the traditional and the SG quantum canonical ensembles}
\label{sec:trad-new-quant}

\subsection{Description of the ensembles}
\label{sec:two-canon-ensembl}

Consider a physical system on a Hilbert space ${\cal H}$ of dimension $n$. We 
define the dynamics by a Hamiltonian  $\hat H$. In this context, the 
traditional quantum canonical ensemble (QCE, see standard textbooks such as
Ref.~\cite{von1955mathematical,Huang1987,reichl1998modern,callen2006thermodynamics} for  details) is the ensemble defined by the density matrix
\begin{equation}
  \label{eq:36}
 \hat  \rho_{C}=Z^{-1}\exp (-\beta \hat H),
\end{equation}
where the partition function $Z$ is defined as
\begin{equation}
  \label{eq:37}
  Z=\mathrm{Tr}\left ( \exp (-\beta \hat H)\right ).
\end{equation}

One may alternatively arrive at ensemble definitions by establishing first a
probability distribution $F(|\psi\rangle)$ on the Hilbert space ${\cal H}$ with
respect to the canonical volume element $d\mu (|\psi\rangle)$, or equivalently,
a probability distribution $F(P_{\psi})$ on the projective space ${\cal PH}$,
where $P_{\psi}$ is the  orthogonal projector onto the one-dimensional subspace generated by state $\psi$. These
distributions can then be used to define the ensembles, i.e., to
construct the 
density matrices: Given any probability distribution $F$ defined on
the state
space, the corresponding density matrix is given by:
\begin{equation}
\label{equation3}
  \hat \rho=\int_{{\cal PH}}d\mu(P_{\psi})F(P_{\psi})P_{\psi},
\end{equation} 
where $d\mu$ represents a (dynamically) 
invariant measure on ${\cal PH}$. Different 
distributions $F$ and $F'$ may lead to the same density matrix, and must then
be considered equivalent. This is just reflecting the well known ambiguity
in the definition of a density matrix: Given a density matrix $\hat \rho$,
there are infinitely many ways of writing it as a convex combination of
rank-one projectors.

One may now prove that the average value of an arbitrary quantum observable 
$\hat O$, can be calculated as (see \cite{Alonso2012a})
\begin{equation}
  \label{eq:5}
  \langle \hat O\rangle=\int_{{\cal PH}}d\mu(P_{\psi})F(P_{\psi})
O(P_{\psi}),
\end{equation}
where $O(P_{\psi})=\mathrm{Tr}(\hat O P_{\psi})=\frac{\langle
  \psi|\hat O\psi\rangle}{\langle\psi|\psi\rangle}$. However the result is exactly the same if we compute
\begin{equation}
  \label{eq:38}
  \langle \hat O\rangle =\mathrm{Tr} (\hat \rho \hat O),
\end{equation}
where $\hat \rho$ is Eq.~(\ref{equation3}).

One can construct one of such probability distributions to obtain the 
quantum canonical ensemble $\hat{\rho}_C$. Let us begin with a
distribution defined on the Hilbert space as: 
\begin{equation}
  \label{eq:41}
  F(|\psi\rangle)=Z^{-1}\sum_{k}e^{-\beta E_{k}}\delta \left (
    |\psi\rangle-|E_{k}\rangle \right
  )=-\sum_{k}\beta^{-1}\frac{\partial \log Z}{\partial E_{k}} \delta \left (
    |\psi\rangle-|E_{k}\rangle \right
  ),
\end{equation}
with the partition function
\begin{equation}
  \label{eq:42}
  Z=\int_{{\cal H}}d\mu(|\psi\rangle)\sum_{k}e^{-\beta E_{k}}\delta \left (
    |\psi\rangle-|E_{k}\rangle \right ) =\sum_{k}e^{-\beta E_{k}},
\end{equation}
where $\{ |E_{k}\rangle\}_{k=1, \cdots, n}$ represent the energy
eigenvectors of the Hamiltonian $\hat H$ and $\{ E_{k}\}$ the
corresponding eigenvalues.  Of course, as we mentioned above, this
representation is not unique.

If we prefer to consider the distribution as defined on the projective space 
${\cal PH}$, we have 
\begin{equation}
  \label{eq:39}
  F(P_{\psi})=Z^{-1}\sum_{k}e^{-\beta E_{k}}\delta \left (P_{\psi}-\frac{|E_{k}\rangle\langle
    E_{k}|}{\langle E_{k}|E_{k}\rangle} \right ),
\end{equation}
where now the partition function is written as:
\begin{equation}
  \label{eq:40}
  Z=\int_{{\cal PH}}d\mu(P_{\psi})\sum_{k}e^{-\beta E_{k}}\delta \left (P_{\psi}-\frac{|E_{k}\rangle\langle
    E_{k}|}{\langle E_{k}|E_{k}\rangle} \right ) =\sum_{k}e^{-\beta E_{k}},
\end{equation}
where $d\mu$ represents an invariant measure on ${\cal PH}$. 

Notice that these are just the expressions for the spectral decomposition of
the density matrix (\ref{eq:36}) written as a probability distribution on ${\cal
H}$ or ${\cal PH}$. These distributions select, out of all states of ${\cal H}$
(equivalently of ${\cal PH}$), only those that are eigenstates of the
Hamiltonian, and assigns to them the corresponding Boltzmann probability. This
may lead one to think that the canonical ensemble contains some kind of
preference for those states, but this is not true. Invoking the above-mentioned
non-uniqueness, one may use a different distribution $F$ to construct the
canonical ensemble $\hat{\rho}_C$, without using delta functions centered at
the Hamiltonian eigenstates. It could be a continuous distribution over the
entire space, or even one distribution that assigns zero probability to those
eigenstates.

Brody and coauthors (see Refs. \cite{Brody1998b,Brody2007b}) and Jona-Lasinio
and Presilla (see Refs. \cite{jona2000invariant,jona2006statistics}) introduced
another distribution, following the seminal idea by Schrödinger
\cite{schrodinger1989statistical}. The main point is to consider, instead of
only one state for each eigenvalue of the Hamiltonian, all physical states of
${\cal H}$ (or ${\cal PH}$) leading to the same expectation value for $\hat H$,
and assign to all of them the Boltzmann probability with respect to that
expectation value, i.e., $(1/Z)\exp(-\beta \langle \psi | \hat{H} | \psi
\rangle)$. As we said, we will call this distribution the Schrödinger-Gibbs
distribution and represent it as $F^{SG}$. Notice that, by construction,
$F^{SG}$ can be thought as a direct generalization to the quantum realm of the
classical canonical distribution. However, as we are going to see, its properties
are different from the classical case because of the quantum nature of the states
it is defined on.

The SG distribution on the Hilbert space can thus be written as:

\begin{equation}
  \label{eq:43}
  F^{SG}(|\psi\rangle)=(Z^{SG})^{-1}\delta(
  1-\langle\psi|\psi\rangle )e^{-\beta \langle \psi |\hat
    H\psi\rangle},
\end{equation}
where now the partition function reads

\begin{equation}
  \label{eq:44}
  Z^{SG}=\int_{{\cal H}}d\mu(|\psi\rangle) \delta( 1-\langle\psi|\psi\rangle
  ) e^{-\beta \langle \psi |\hat
    H\psi\rangle}= \int_{S}d\mu(|\psi\rangle)e^{-\beta \langle \psi 
    |\hat H\psi\rangle} ,
\end{equation}
where we denote by $S$ the sphere of vectors in ${\cal H}$ of norm equal to
one \footnote{There is an ambiguity in the definition of the volume element on
the sphere by using the constraint, but it represents just a global
multiplicative constant, which is irrelevant to define the corresponding
statistical system.}. This last distribution is equivalent to a distribution
$F^{SG}$ defined on the projective space ${\cal PH}$ modulo a constant
factor that does not affect the resulting ensemble (if interested see
Appendix \ref{sec:integr-proj-space} to understand the origin of the constant
factor). 
If we define  the expectation value of the operator
$\hat H$ on projector
$P_{\psi}=\frac{|\psi\rangle\langle\psi|}{\langle\psi |\psi\rangle}\in {\cal PH}$ as
\begin{equation}
  \label{eq:47}
   H(P_{\psi}):=\mathrm{Tr}(P_{\psi}\hat H),
\end{equation}
the distribution can be written as:

\begin{equation}
  \label{eq:45}
  F^{SG}(P_{\psi})=Z_{SG}^{-1}e^{-\beta H(P_{\psi})},
\end{equation}
with partition function
\begin{equation}
  \label{eq:46}
 Z^{SG}=\int_{{\cal PH}}d\mu(P_{\psi})e^{-\beta H(P_{\psi})}. 
\end{equation}

If we evaluate the density matrix associated with the distribution (\ref{eq:45}) (and
use the result proved in Appendix \ref{sec:integr-proj-space} to evaluate the integrals
on the sphere $S$ of vectors of ${\cal H}$ with norm equal to one) we obtain:
\begin{equation}
  \label{eq:49}
  \hat \rho^{SG}=(Z^{SG})^{-1}\int_{S}d\mu (|\psi\rangle)e^{-\beta
    \langle \psi| \hat H\psi\rangle}|\psi\rangle\langle\psi|.
\end{equation}

Regarding the measure $d\mu$ a few comments are in order.
It is well known (see \cite{Alonso2011,Alonso2012a} and references
therein) that finite dimensional quantum systems defined on
$\mathbb{C}^{n}$ admit a Hamiltonian structure associated with the
canonical Kähler structure of the manifold. The  Schrödinger equation
corresponds then to the flow of a Hamiltonian vector field associated
with the canonical symplectic form. Therefore, the Liouville theorem ensures
that the corresponding symplectic volume is preserved by the dynamics, as it happens
in Classical Mechanics.   As the sphere of vectors with norm one is
also preserved by the dynamics, the natural choice for $d\mu$ on  $S$
is the restriction of the symplectic volume of $\mathbb{C}^{n}$.
Regarding  the projective space ${\cal PH}$, again there exists a
canonical Kähler structure on it, also preserved by the
dynamics. Therefore we can also consider the corresponding symplectic
volume, which will also be invariant. 

Now, what the physical origin of the differences between Eq.~(\ref{eq:36})
and Eq.~(\ref{eq:49}) is. If one approaches the canonical ensemble from the
microcanonical one the origin is clear. Indeed, while the microcanonical in
Refs.~\cite{Brody1998b,Brody2007b,Bender2005a} is based on the postulate that
every state possessing the same energy expectation value must be democratically
realized (which implements Schrödinger's suggestion), in the traditional
microcanonical ensemble, orthogonal states must be the only ones to be included
in the sample space of the statistical thermodynamics. The arguments for using
only orthogonal states are based on the standard theory of probability (see for
instance Secs~4.5 and 4.6 of Ref.~\cite{griffiths2003consistent} and Ref
\cite{Hohenberg2010}), and on the relationship between information theory and
thermodynamics
\cite{Jaynes1957,Fuchs1995,Cafaro2013,nielsen2010quantum,peres1995quantum,Maruyama2009,Kim2011,Plesch2013,Kim2013}.

\subsection{Quantum properties of the Schrödinger-Gibbs distribution}
\label{sec:quant-prop-boltzm}

Let us now study the properties of the density matrix (\ref{eq:49}) in more
detail.
First of all, for technical reasons that will become clear later, we want to 
prove that the SG density matrix commutes with the Hamiltonian and hence both
operators can be diagonalized in a common eigenbasis.
Notice that if we consider $\hat \rho^{SG}$ as a stationary distribution
corresponding to an equilibrium situation and the Hamiltonian does not
depend on time, the result is trivial:
\begin{equation}
  \label{eq:52}
  i\hbar\frac{\partial \hat \rho^{SG}(t)}{\partial t}=[\hat H, \hat
  \rho^{SG}(t)]\quad \longrightarrow \quad \frac{\partial \hat
    \rho^{SG}(t)}{\partial t}=0 \Rightarrow [\hat H, \hat
  \rho^{SG}(t)]=0.
\end{equation}

However \textit{a priori} there is no reason to consider $\hat{\rho}^{SG}$  to be
stationary. In the
following, we will prove that it is indeed time-independent. 

 The simplest way to do
this is to consider a unitary evolution for $\hat\rho^{SG}$ of the form
\begin{equation}
  \label{eq:33}
 \hat  \rho^{SG}(t)=\hat U(t)\hat \rho^{SG} (0) \hat U^{\dagger}(t),
\end{equation}
where, as we assume that the Hamiltonian does not depend on time, 
\color{black}
\begin{equation}
  \label{eq:64}
  \hat U(t):=e^{-i\hat H t}.
\end{equation}
\color{black}

Under the same evolution, the volume element $d\mu$ on $S$ and the expectation
value of the energy $\langle \psi| \hat H \psi\rangle$ are, because of
unitarity, constant in ``time''. Thus, for any $t$, we can write
$$
\langle \psi (t)| \hat H \psi (t)\rangle=\langle \psi (0)| \hat H \psi
(0)\rangle  \qquad \forall t,
$$
and thus, as the volume element is also preserved, we find
$$ 
d\mu (|\psi(0)\rangle)Z_{SG}^{-1}e^{-\beta \langle \psi(0)|
   \hat H\psi(0)\rangle}=
 d\mu (|\psi(t)\rangle)Z_{SG}^{-1}e^{-\beta \langle \psi(t)|
   \hat H\psi(t)\rangle}\qquad \forall t.
$$

\color{black}
Consider now the expression of the density matrix $\hat \rho^{SG}$
defined in Eq. (\ref{eq:49}) which we denote  as $\hat
\rho^{SG}_{0}$.  Let us
consider how  the unitary evolution transforms this density
matrix after  an arbitrary but fixed value of time $t_{*}$, i.e.,
\color{black}
\begin{equation}
  \label{eq:55}
\hat   \rho^{SG}(t_{*})=\hat U(t_{*})\hat \rho^{SG}_{0}\hat U^{\dagger}(t_{*})=
 Z_{SG}^{-1} \int_{S}d\mu (|\psi\rangle)e^{-\beta \langle \psi|  
 \hat H \psi \rangle}  \hat U(t_{*})|\psi\rangle\langle\psi | \hat
 U(t_{*})^{\dagger}.
\end{equation}
Here $\hat U(t_{*})$ defines a transformation
on the set of states as 
\begin{equation}
  \label{eq:54}
  |\psi'\rangle=\hat U(t_{*})|\psi\rangle,
\end{equation}
 which obviously preserves the sphere $S$ since $\langle
\psi|\psi\rangle=\langle \psi'|\psi'\rangle$ and thus we
can re-write Eq. (\ref{eq:55}) as  
\begin{align}
  \label{eq:71}
 \hat  \rho^{SG}(t_{*}) =Z_{SG}^{-1}\int_{S}d\mu (|\psi\rangle)e^{-\beta \langle \psi|
   \hat H\psi\rangle}|\psi'\rangle\langle\psi'|.
\end{align}
We know that the measure is invariant, i.e., 
\begin{equation}
  \label{eq:56}
  d\mu (|\psi\rangle)Z_{SG}^{-1}e^{-\beta \langle \psi|
   \hat H\psi\rangle}=d\mu (|\psi'\rangle)Z_{SG}^{-1}e^{-\beta \langle \psi'|
   \hat H\psi'\rangle},
\end{equation}
and therefore,  as the sphere $S$ is invariant and the
integral runs over all normalized states, we can write 
\begin{align}
  \label{eq:71}
 \hat  \rho^{SG}(t_{*}) =Z_{SG}^{-1}\int_{S}d\mu (|\psi'\rangle)e^{-\beta \langle \psi'|
   \hat H\psi'\rangle}|\psi'\rangle\langle\psi'|,
\end{align}
\color{black}
which is the same as Eq. (\ref{eq:49}). Hence, we have
proved that 
\begin{equation}
  \label{eq:57}
  \hat \rho^{SG}(t_{*})=\hat \rho^{SG}_{0}.
\end{equation}

As the value of $t_{*}$ was generic, we conclude that the  density
matrix is independent of $t$ and hence  commutes with the Hamiltonian:
\begin{equation}
  \label{eq:25}
  [\hat H, \hat \rho^{SG}]=0.
\end{equation}

\color{black}

On a different matter, in order to compute the integral in
eq.~(\ref{eq:49}), we can use the tools of complex analysis, as  presented
in Appendix \ref{sec:partition-function}. The result for the partition function is
obtained in Eqs.~(\ref{eq:76}-\ref{eq:80}). Assuming that there exist $p+1$
different eigenvalues of the Hamiltonian $\hat H$ with degeneracies $d_{0},
\cdots, d_{p}$, the partition function is written as:
\begin{equation}
  \label{eq:76b}
  Z^{SG}=\sum_{k=0}^{p}e^{-\beta E_{k}}F_{E_k},
\end{equation}
where 
\color{black}
\begin{equation}
  \label{eq:80b}
  F_{E_k}=-\frac {(2\pi)^{n}}{(d_{k}-1)!}\sum_{\substack{j_{0}, \cdots,
    j_{p}= 0\\ j_{0}+j_{1}+\cdots
    j_{p}=d_{k}-1}}^{d_{k}-1}
\binom{d_{k}-1}{j_{0},\cdots, j_{p}}
\prod_{\substack{s=
  0\\ s\neq k}}^{p}\frac{(-1)^{j_{0 }}\frac{(d_{s}+j_{s}-1)!}{(d_{s}-1)!}}{(\beta E_{s}-\beta E_{k})^{d_{s}+j_{s}}},
\end{equation}
\color{black}
and $\binom{d_{k}-1}{j_{0}, \cdots, j_{s}}$ represents the multinomial 
coefficient
$$
\binom{d_{k}-1}{j_{0}, \cdots, j_{s}}=\frac{(d_{k}-1)!}{j_{0}! \cdots j_{s}!}. 
$$

Notice that the traditional distribution (\ref{eq:41}) associated with the QCE 
corresponds to the case where 
\begin{equation}
  \label{eq:79}
  F_{E_{k}}=d_{k}; \qquad \forall k=0, \cdots, p,
\end{equation}
where $d_{k}$ stands for the degeneracy of the energy level $E_{k}$.

From the above expressions, we can obtain the spectral decomposition of the
density matrix $\hat \rho^{SG}$ defined by Eq. (\ref{eq:49}). Indeed, we
know that $\hat \rho^{SG}$ is diagonal in the energy-eigenbasis, and
thus it can be written as
\begin{equation}
  \label{eq:8}
  \hat \rho^{SG}=\sum_{k} \omega_{k}\frac{|E_{k}\rangle\langle
  E_{k}|}{\langle E_{k}|E_{k}\rangle },
\end{equation}
where the eigenvalue $\omega_{k}$ corresponding  to the eigenvector
$|E_{k}\rangle$ is equal to: 

\begin{equation}
  \label{eq:48}
  \omega_{k}= \hat \rho^{SG}_{kk}=Z_{SG}^{-1}\int_{S}d\mu(|\psi\rangle)e^{-\beta
    \sum_{j }E_{j}|\psi_{j}|^{2}} |\psi_{k}|^{2},
\end{equation}
and $|\psi\rangle=\sum_{j }\psi_{j}|E_{j}\rangle$. 
Each $\omega_{k}$ is thus obtained as a suitable derivative of the
partition function $Z^{SG}$, as  in the case of the
canonical distribution:
\begin{equation}
  \label{eq:84}
  \omega_{k}=-\beta^{-1}\frac{\partial \log Z^{SG}}{\partial E_{k}}=
  \frac{e^{-\beta E_{k}}F_{E_{k}}}{Z^{SG}}
  -\beta^{-1}\frac{\sum_{j= 0}^{p}e^{-\beta E_{j}}\frac{\partial
      F_{E_{j}}}{\partial E_{k}}}{Z^{SG}}.
\end{equation}
Therefore, we just proved that the probability distribution (\ref{eq:43}) can be written 
as
\begin{equation}
  \label{eq:59}
  F^{SG}(|\psi\rangle)=\sum_{k} \left ( \frac{e^{-\beta
      E_{k}}F_{E_{k}}}{Z^{SG}}-
                    \beta^{-1}\frac{\sum_{j= 0}^{p}e^{-\beta E_{j}}\frac{\partial
      F_{E_{j}}}{\partial E_{k}}}{Z^{SG}} \right ) \delta(|\psi\rangle-|E_{k}\rangle),
\end{equation}
which is formally analogous to the distribution (\ref{eq:41}), but with very 
different content because of the second term in the sum. If we write it as a 
density matrix we have:
\begin{equation}
  \label{eq:85}
  \hat \rho^{SG}=\sum_{k} \left ( \frac{e^{-\beta
      E_{k}}F_{E_{k}}}{Z^{SG}}-
                    \beta^{-1}\frac{\sum_{j= 0}^{p}e^{-\beta E_{j}}\frac{\partial
      F_{E_{j}}}{\partial E_{k}}}{Z^{SG}} \right ) \frac{|E_{k}\rangle\langle
  E_{k}|}{\langle E_{k}|E_{k}\rangle }.
\end{equation}
Notice that the second term is such that $\sum_k
\frac{\partial F_{E_j}}{\partial E_k}=0$ and then $\mathrm{Tr} \hat \rho^{SG}=1$.

Only if the functions $F_{E_{k}}$ were independent from the energy eigenvalues and
equal to the degeneracies $d_{k}$,  would $\hat \rho^{SG}$ coincide with the
usual canonical distribution $\hat \rho_{C}$ in Eq.~(\ref{eq:36}). Thus, given
any $\beta$, the average value of any quantum observable $\hat O$ will be in general
different in the  ensembles 
and, in general, the properties of the  ensembles will not
coincide. We will return to this point in Section \ref{sec:additivity-entropy}.

A similar computation  for some particular cases  can be found in
\cite{Brody2010c}. Indeed, the authors analyze the completely
non-degenerate case of a  general system, and the strong
coupling limit of a spin-spin interaction Hamiltonian. Nonetheless,
the authors of \cite{Brody2010c} do not study the implications of
their results at the thermodynamic level, as we will in the next
section. 

\subsection{Thermodynamical properties of the Schrödinger-Gibbs distribution}
\label{sec:therm-prop-boltzm}

We have just seen how, even if $F^{SG}$ is analogous to the canonical
distribution of a classical (continuous) system, the quantum nature of the
system it describes makes it equivalent to a quantum ensemble described by a
density matrix $\hat \rho^{SG}$ that is different from the traditional 
canonical one. We want to check now if, despite these differences, we can still 
define an appropriate thermodynamics associated with the SG distribution.

In order to define a system in thermodynamical equilibrium, we must obtain the 
corresponding thermodynamic magnitudes, both extensive and intensive ones, from 
the statistical-mechanics objects. 
Thus, for a system described by a partition function $Z$,  the Helmholtz
free energy $F$ (which is the natural free energy in the  
canonical ensemble) corresponds to
\begin{equation}
  \label{eq:86}
 F=-\beta^{-1}\log Z,
\end{equation}
the thermodynamic (also called sometimes ``internal'') energy $U$ corresponds 
to the average value of the energy, which can be obtained through
\begin{equation}
  \label{eq:87}
  U=-\frac{\partial \log Z}{\partial \beta},
\end{equation}
and the thermodynamic entropy $S^{th}$ is written as
\begin{equation}
  \label{eq:88}
  S^{th}=k_{B}\left (\log Z-\beta \frac{\partial \log Z}{\partial \beta} \right ).
\end{equation}
If the partition function $Z$ is factorizable, the functions $F, U$ and
$S^{th}$ are extensive magnitudes.
Then, the extensiveness of the  functions $F$, $U$ or $S^{th}$ in the present
case, is equivalent to the factorizability property of the partition function 
$Z^{SG}$. That is, if we consider a system defined as the composition of 
non-interacting subsystems, the partition function of the complete system must 
be written as the product of the partition function restricted to the subsystems 
in order to yield extensive thermodynamic functions. Notice that, so far, $\beta$ 
represents just a parameter that is used to define the partition function and ultimately
may not have any physical meaning  (for example, if the resulting
thermodynamics is found to be inappropriate).

Consider then that the total Hilbert space ${\cal H}$ is equal to the tensor 
product of the Hilbert spaces describing the corresponding subsystems, i.e.,
\begin{equation}
  \label{eq:89}
  {\cal H}={\cal H}_{1}\otimes \cdots \otimes {\cal H}_{m},
\end{equation}
with dimensions $n_{1}, n_{2}, \cdots, n_{m}$ (for simplicity we
assume the case of finite-dimensional quantum systems). The dimension
of ${\cal H}$ is thus given by $n=n_{1}\cdots n_{m}$. Since we assume
that there is no interaction among the subsystems, the total Hamiltonian is 
written as the sum
\begin{equation}
  \label{eq:90}
  \hat H=\sum_{\alpha=1}^{m}\hat H_{\alpha}=\sum_{\alpha=1}^{m}  \hat{\mathbb{I}}_{n_{1}}\otimes\cdots\otimes 
  \hat{\mathbb{I}}_{n_{\alpha-1}}\otimes \hat{h}_{\alpha}\otimes \hat{\mathbb{I}}_{n_{\alpha+1}}
  \otimes \cdots\otimes \hat{\mathbb{I}}_{n_{m}}. 
\end{equation}

As all $\hat H_{\alpha}$ commute, the tensor product of the eigenbases of the
$\hat h_{\alpha}$ is an eigenbasis of the total Hamiltonian $\hat{H}$. We thus have 
that each element of the total-energy eigenbasis $\{
|E_{k}\rangle\}$(with corresponding eigenvalues $E_k$) can be  
written as the tensor product of the corresponding elements of the different 
eigenbases $\{|e^{\alpha}_{j_{\alpha}}\rangle \}$ of the subsystems (with corresponding eigenvalues $e^\alpha_k$) :
\begin{equation}
  \label{eq:91}
  |E_{k}\rangle=|e_{j_{1}}^{1}\rangle\otimes \cdots \otimes
  |e_{j_{m}}^{m}\rangle; \qquad E_{k}=e_{j_{1}}^{1}+\cdots + e_{j_{m}}^{m}.
\end{equation}
With no loss of generality, we will assume that
\begin{equation}
  \label{eq:93}
  E_{k}\leq E_{k+1}; \qquad e^{\alpha}_{j_{\alpha}}\leq
  e^{\alpha}_{j_{\alpha}+1},\qquad \forall \alpha.
\end{equation}

It is well known that the partition function of the canonical distribution
$\hat \rho_{C}$ in Eq.~(\ref{eq:36}) factorizes as a product of the partition
function of the subsystems (see \cite{callen2006thermodynamics}, Sec.~16.2). If
we compute the partition function $Z^{SG}$ in this basis, we obtain as a result
Eq.~(\ref{eq:76b}). If we did the same computation restricted to the subsystem
$\alpha$, we would obtain similar expressions for the partition functions
$Z^{SG}_{\alpha}$, with the eigenvalues $E_{k}$ replaced by the eigenvalues
$e^{\alpha}_{j_{\alpha}}$. In order to check whether or not $Z^{SG}$
factorizes, let us consider the limit  $\beta>>1$, where we can write that
\begin{equation}
  \label{eq:92}
  Z^{SG}|_{\beta>>1}\sim e^{-\beta E_{0}}F_{E_{0}}. 
\end{equation}
Analogously, 
\begin{equation}
  \label{eq:94}
  Z^{SG}_{\alpha}|_{\beta>>1}\sim e^{-\beta e^{\alpha}_{0}}F_{e^{\alpha}_{0}}. 
\end{equation}

Factorizability of the partition function would require that
\begin{equation}
 \left . Z^{SG} \right |_{\beta>>1}=\left .  \prod_{\alpha=1}^{m} Z^{SG}_\alpha \right |_{\beta>>1},
\end{equation}
which implies:
\begin{equation}
   F_{E_{0}}= \prod_{\alpha=1}^{m}F_{e_{0}^{\alpha}}.
\end{equation}

However  we can immediately check that \footnote{For simplicity we will consider the
case of non-degenerate ground states.}
\begin{equation}
  \label{eq:95}
  F_{E_{0}}\neq \prod_{\alpha=1}^{m}F_{e_{0}^{\alpha}}.
\end{equation}
Indeed, in the non-degenerate  groundstate  case,  Eq. (\ref{eq:80b}) becomes
for $E_k=E_0$  and $E_k=e_0^\alpha$ , respectively 
\begin{equation}
  \label{eq:96}
  F_{E_{0}}=\prod_{j\neq 0} \frac{1}{\beta E_{j}-\beta E_{0}}; \qquad
  F_{e_{0}^{\alpha}}=\prod_{j_{\alpha}\neq 0}\frac{1}{\beta
    e^{\alpha}_{j_{\alpha}}-\beta e^{\alpha}_{0}},
\end{equation}
and thus it is simple to prove that
\begin{equation}
  \label{eq:97}
  \prod_{j\neq 0} \frac{\beta^{-1}}{E_{j}-E_{0}}\neq
  \prod_{\alpha}\prod_{j_{\alpha}\neq 0}\frac{\beta^{-1}}{e^{\alpha}_{j_{\alpha}}-e^{\alpha}_{0}}.
\end{equation}

The last inequality can be  easily understood  by thinking that, on the right hand side, 
there are only energy differences corresponding to a change in one subsystem, 
while, on the left hand side, all possible energy differences are considered.

This being so in the limit $\beta>>1$   is sufficient to prove that the
partition function $Z^{SG}$ is not the product of the partition
functions of the subsystems, and therefore that we cannot define extensive
magnitudes from $Z^{SG}$. Besides, this property does not depend on the number
of subsystems, and therefore it is still valid in a thermodynamic limit where
$m\to \infty$; i.e., we can claim that
\begin{equation}
  \label{eq:50}
  Z^{SG}\neq \prod_{\alpha=1}^{\infty}Z_{\alpha}^{SG}.
\end{equation}

We thus conclude that the thermodynamic functions defined by
Eqs.~(\ref{eq:86}), (\ref{eq:87}) or~(\ref{eq:88}) cannot represent extensive 
magnitudes.

In principle we may study the same problem from the point of view of the
density matrix $\hat \rho^{SG}$, but doing so in full generality becomes quite
difficult from the computational point of view. We will tackle this analysis in
Sec.~\ref{sec:additivity-entropy} for the case of a specific quantum system
resulting from the composition of two-level systems, and we will recover the 
same results obtained here but in a much simpler way.

\subsection{The origin of non-extensiveness: integrating over entangled states}
\label{sec:orig-non-extens}

Let us consider the simplest case of a composite system by assuming that we have
$m=2$ in Eqs.~(\ref{eq:89}) and (\ref{eq:90}).  We know from
Eq. (\ref{eq:50}) that 
\begin{equation}
  \label{eq:16}
  Z_{1}^{SG}Z_{2}^{SG}\neq Z_{12}^{SG}.
\end{equation}
Now, we want to understand the origin of this difference from a physical
point of view.
 
The partition functions for the
individual subsystems and for the composite system will read, respectively:
\begin{equation}
  \label{eq:98}
  Z_{1}^{SG}=\int_{{\cal S}^{n_{1}}}d\mu(|\psi^{1}\rangle)e^{-\beta \langle \psi^{1} | \hat H_{1}
  \psi^{1}\rangle},
\end{equation}
\begin{equation}
  \label{eq:98b}
  Z_{2}^{SG}=\int_{{\cal S}^{n_{2}}}d\mu(|\psi^{2}\rangle)e^{-\beta \langle \psi^{2} | \hat H_{2}
  \psi^{2}\rangle},
\end{equation}
and
\begin{equation}
  \label{eq:98c}
  Z_{12}^{SG}=\int_{{\cal S}^{n_{1}n_{2}}}d\mu(|\psi\rangle)e^{-\beta \langle \psi | \hat H
  \psi\rangle},
\end{equation}
where ${\cal S}^{n_{1}}, {\cal S}^{n_{2}}$ and ${\cal S}^{n_{1}n_{2}}$ represent the
$(n_1-1)$-dimensional, $(n_{2}-1)$--dimensional and
$(n_{1}n_{2}-1)$-dimensional spheres.

We can write the product of two integrals like that of Eq.~(\ref{eq:98}) as
\begin{equation}
  \label{eq:13}
( Z_{1}^{SG})( Z_{2}^{SG})=  \int_{{\cal S}^{n_{1}}\times
  {\cal S}^{n_{2}}}d\mu_{1}(|\psi^{1}\rangle)d\mu_{2}(|\psi^{2}\rangle)  e^{-\beta \langle \psi ^{1}| \hat H_{1}
  \psi^{1}\rangle} e^{-\beta \langle \psi ^{2}| \hat H_{2}
  \psi^{2}\rangle}.
\end{equation}

In a basis like the one in~(\ref{eq:91}), constituted by separable vectors, we 
have
\color{black}
\begin{equation}
  \label{eq:9}
  e^{-\beta \langle \psi ^{1}| \hat H_{1}
  \psi^{1}\rangle} e^{-\beta \langle \psi ^{2}| \hat H_{2}
  \psi^{2}\rangle}=e^{-\beta\sum_{k= 1}^{n_{1}}
  e^{1}_{k}|\psi^{1}_{k}|^{2}}e^{-\beta\sum_{j= 1}^{n_{2}}
  e^{2}_{j}|\psi^{2}_{j}|^{2}}  \ .
\end{equation}
The separability of the vectors allows us to write 

\begin{equation}
  \label{eq:12}
    \sum_{k=1}^{n_{1}}e^{1}_{k}|\psi^{1}_{k}|^{2}+\sum_{j=1}^{n_{2}}e^{2}_{j}|\psi^{2}_{j}|^{2}=\langle
  \psi |\hat H \psi\rangle,\qquad \psi=\psi^{1}\otimes \psi^{2}.
\end{equation}
\color{black}
Thus  the resulting exponent is the expectation value of the Hamiltonian $\hat 
H=\hat H_{1}\otimes \mathbb{I}_{2}+\mathbb{I}_{1}\otimes \hat H_{2}$, evaluated at the separable points of ${\cal 
H}={\cal H}_{1}\otimes {\cal H}_{2}$.  Notice that in the sum above,
$k$ runs from 1 to $n_{1}$ and $j$ from 1 to $n_{2}$, in such a
way that the coordinate expression of a separable state $\psi$ reads
\begin{equation}
  \label{eq:51}
  \psi=\psi^{1}\otimes \psi^{2}=\left (\sum_{k}
    \psi^{1}_{k}|e^{1}_{k}\rangle \right ) \otimes \left (\sum_{j}
    \psi^{2}_{j}|e^{2}_{j}\rangle \right ).
\end{equation}
Putting everything together, we have just proved that:
\begin{equation}
  \label{eq:15}
  ( Z_{1}^{SG})( Z_{2}^{SG})=\int_{{\cal S}^{n_{1}}\times
  {\cal S}^{n_{2}}}d\mu_{1}(|\psi^{1}\rangle)d\mu_{2}(|\psi^{2}\rangle)
e^{-\beta \langle
  \psi |\hat H \psi\rangle} \ .
\end{equation}
Equation (\ref{eq:15}) sums the same function
as (\ref{eq:98c}), but only over the separable states of ${\cal
S}^{n_{1}n_{2}}\subset {\cal H}={\cal H}_{1}\otimes {\cal H}_{2}$; while
(\ref{eq:98c}) integrates over all states, both separable and
entangled. This explains  Eq. (\ref{eq:16}) from the physical point of
view.

If we compare this situation with the traditional canonical distribution
described by Eq.~(\ref{eq:41}), we see that the delta functions would restrict
the corresponding integrals to the eigenvectors of the Hamiltonian $\hat H$,
which, in the case of non-interacting subsystems, correspond always to
separable states. In addition, the statistical weight assigned to each of these
eigenstates depends only on the eigenspace. In contrast, if we think about
the SG distribution in terms of the corresponding density matrix (i.e., using
expression~(\ref{eq:85})), we would be considering, apparently, only separable
states (the Hamiltonian eigenstates participating of the spectral
decomposition), but now the statistical weights associated with them 
do not factorize. Therefore  the density matrix (\ref{eq:85})
is not separable.  We will illustrate
this in a simple example in next section.

\section{A simple example}
\label{sec:additivity-entropy}

In this section we will present numerical examples of the properties discussed 
in the previous section. We will analyze the two different distributions 
((\ref{eq:41}) and (\ref{eq:44})) for a system of $N$ noninteracting two-level
particles whose dynamics is described by a Hamiltonian of the form
\begin{equation}
  \label{eq:6}
  \hat H=\sum_{k}\hat{H}_{k}=\sum_{k}  \overbrace{\hat{\mathbb{I}}_{2}\otimes 
 \cdots\otimes \hat{\mathbb{I}}_{2}}^{k-1}\otimes \hat{h}_{k}\otimes   \overbrace{\hat{\mathbb{I}}_{2}\otimes
\cdots\otimes\hat{\mathbb{I}}_{2} }^{N-k}\, ,
\end{equation}
where $\hat{\mathbb{I}}_{2}$ stands for the identity operator in two
dimensions and the one-particle Hamiltonian $\hat{h}_{k}$ can be written in
the corresponding eigenbasis as
\begin{equation}
  \label{eq:7}
  h_{k}=
  \begin{pmatrix}
    0 & 0 \\
0 & \Delta
  \end{pmatrix},\qquad \forall k=1, \cdots, N,
\end{equation}
where $\Delta$ represents the energy-gap.

Let us now consider a basis for ${\cal H}$ defined as the tensor product of the 
energy eigenbases of every two-level subsystem. Thus, if we write $n=2^{N}$ for short, we have the set of vectors of the form
\begin{equation}
  \label{eq:11}
  |i_{1}, \cdots , i_{n}\rangle=|i_{1}\rangle\otimes \cdots\otimes 
  |i_{n}\rangle; \qquad \text{ where }\ |i_{k}\rangle=
  \begin{cases}
    |0\rangle \\
|1\rangle
  \end{cases}.
\end{equation}
We also fix an ordering for the basis:
\begin{equation}
\label{eq:26}
{\cal B}=\{  |0,0, \cdots, 0,0\rangle,\,\, |0, 0, \cdots, 0, 1\rangle, \,\,
|0, 0, \cdots, 1, 0\rangle, \cdots,\,\ ,|0, 1,
  \cdots, 1, 1\rangle, \,\, |1, 1, \cdots, 1\rangle\}.
\end{equation}

In this basis, it is a simple task to verify that the eigenvalues $\{E_{k}\}$ 
of the total Hamiltonian $\hat{H}$ and the corresponding degeneracies $d_{k}$  
are given by 
\begin{equation}
  \label{eq:30}
  E_{k}= k \Delta; \qquad 
  d_{k}=\binom{N}{k},\qquad k=0, \cdots, N.
\end{equation}

\subsection{The partition function and factorizability}

Our goal now is to obtain the expressions for the Schrödinger-Gibbs 
distribution corresponding to this Hamiltonian. We saw that the ensemble is 
defined by the partition function in~(\ref{eq:44}). Thus, our goal is to
write 
\begin{equation}
  \label{eq:27}
  Z^{SG}
=\int_{{\cal S}^{2^{N+1}-1}}d\mu(\psi) e^{-\beta \langle \psi |
      \hat{H}\psi\rangle},
\end{equation}
for the particular case of the Hamiltonian in Eq.~(\ref{eq:6}). Previously, 
from the analysis in Appendix \ref{sec:partition-function}, the expression for $Z^{SG}$ 
was written as in Eq.~(\ref{eq:76b}), where now $n=2^{N}$ and the spectrum is 
given by Eq.~(\ref{eq:30}).

The simplest cases can be easily written. Indeed, we can consider the case 
$N=1$ (i.e., $n=2$) and write Eq.~(\ref{eq:76b}) as
\begin{equation}
  \label{eq:31}
  Z^{SG}_{N=1}=\frac{(2\pi)^{2} \left(1-e^{-\beta  \Delta }\right)}{\beta  \Delta }.
\end{equation}
Analogously, the next case is $N=2$ (or, equivalently, $n=4$):
\begin{equation}
  \label{eq:35}
  Z^{SG}_{N=2}
= \frac{(2\pi) ^4  (1-2\beta \Delta e^{-\beta
      \Delta}-e^{-2\beta \Delta})}{2\beta ^3 \Delta ^3}. 
\end{equation}

We can immediately check that, as we proved in general, the partition function 
does not factorize, i.e.,
\begin{equation}
  \label{eq:63}
  Z^{SG}_{N=2}\neq (Z^{SG}_{N=1})^{2}.
\end{equation}

\subsection{Thermodynamic entropies and specific heat} 

We can also compute the corresponding entropy functions $S^{th}$ using
Eq.~(\ref{eq:88}). We obtain:
\begin{equation}
  \label{eq:65}
  S^{th}_{N=1}=\frac{k_B \left(-\beta  \Delta +e^{\beta  \Delta }+\left(e^{\beta  \Delta }-1\right) \log
   \left(\frac{(2\pi)^{2}  \left(1-e^{-\beta  \Delta }\right)}{ \beta  \Delta
   }\right)-1\right)}{e^{\beta  \Delta }-1},
\end{equation}
and
\color{black}
\begin{align}
 \label{eq:70}
 \notag
S^{th}_{N=2}=&k_{B}\left (
\frac{\left(-2 \beta  \Delta  e^{\beta  \Delta }+e^{2 \beta  \Delta }-1\right) \log \left(\frac{8\pi ^4 e^{-2 \beta  \Delta } \left(-2 \beta  \Delta 
   e^{\beta  \Delta }+e^{2 \beta  \Delta }-1\right)}{ \beta ^3 \Delta
 ^3}\right)}{-2 \beta  \Delta  e^{\beta  \Delta
   }+e^{2 \beta  \Delta }-1} \right .
   \\
&\left .+\frac{-2 \beta  \Delta -2 \beta  \Delta  e^{\beta  \Delta } (\beta  \Delta
   +2)+3 e^{2 \beta  \Delta }-3}
{-2 \beta  \Delta  e^{\beta  \Delta
   }+e^{2 \beta  \Delta }-1}\right ) .
\end{align}
\color{black}

Again, it is simple to verify that these functions are not additive, i.e., that 
\begin{equation}
  \label{eq:72}
  S^{th}_{N=2}\neq 2S^{th}_{N=1}.
\end{equation}

Therefore, we must conclude that $S^{th}$ cannot model an extensive magnitude,
and hence it cannot represent the thermodynamic entropy of a physical
system. Besides this non-extensiveness, $S^{th}$ does not satisfy other
properties that are required for the thermodynamic entropy, such as
positiveness and that it must tend to zero in the limit where
$T=\frac{1}{k_{B}\beta}$ tends to zero. Of course, in this case, $T$ must be
considered just a parameter and in no way can it be identified with a physical
temperature.   Notice that  this behavior does not follow
from the non-extensiveness of the ensemble, since it is present in the
$N=1$ case where non-extensiveness is meaningless. As we mentioned in
the previous Section, the ensemble encoded in Eq (\ref{eq:85}) is
completely different from the canonical ensemble (\ref{eq:36}) and
therefore its properties will, in general, be
different. Thus it is natural that $\hat{\rho}^{C}$ defines extensive
thermodynamical functions while $\hat{\rho}^{SG}$ does not  and it is also
natural that the behavior in the limit $T\to 0$ of the corresponding
entropies differ. Regarding the nonextensiveness of $\hat{\rho}^{SG}$, 
we proved in Section \ref{sec:orig-non-extens} that it is related to
the integration over entangled states in the definition of the
ensemble (Eq. \ref{eq:46}). We do not have a similar proof to explain
the behavior of the entropy in the $T\to 0$ limit, but it is evident
that entanglement is not the only reason.

\begin{figure}[ht]
  \centering
  \includegraphics[width=15cm]{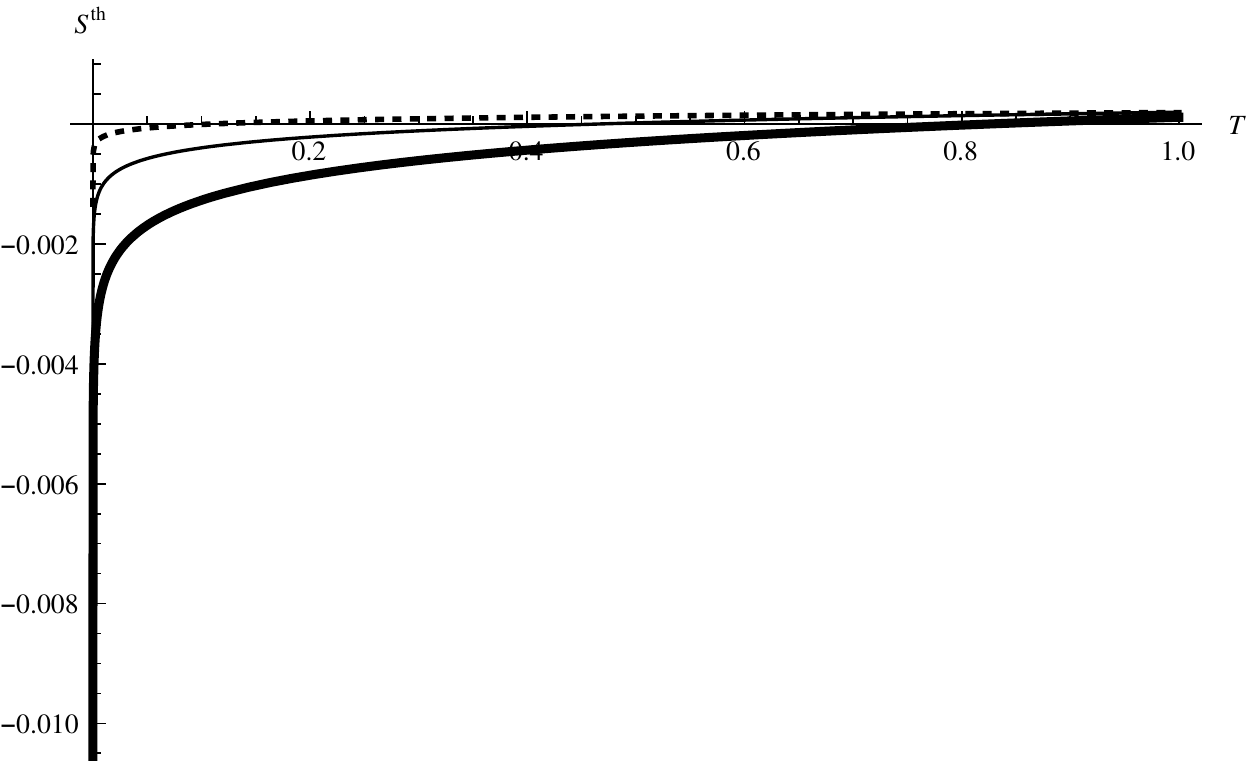}
  \caption{Plot of the functions $S^{th}_{N=1}$ (dotted line), $S^{th}_{N=2}$
    (thin line),  $S^{th}_{{N=3}}$ (thick line) versus T for
    $k_{B}=8.617\times 10^{-5}$ and $\Delta=0.001$ for the SG
    ensemble. We can easily verify the 
    non-additivity property of the function $S^{th}$ and the violation
  of the Third Law, since the entropy does not go to zero in the limit
  $T\to 0$} 
  \label{fig5}
\end{figure}

If we also compute the case $N=3$  (see \footnote{Check the Mathematica notebook included as Supplemental Material at for the functions at any order (results are too long to detail them here)}), we can
represent the corresponding entropies  
together with the previous ones and confirm that, again, their zero-$T$ limit 
is different from zero and the functions are negative in a measurable part of 
their domains (see Fig.~\ref{fig5}). 

\begin{figure}[ht]
  \centering
  \includegraphics[width=15cm]{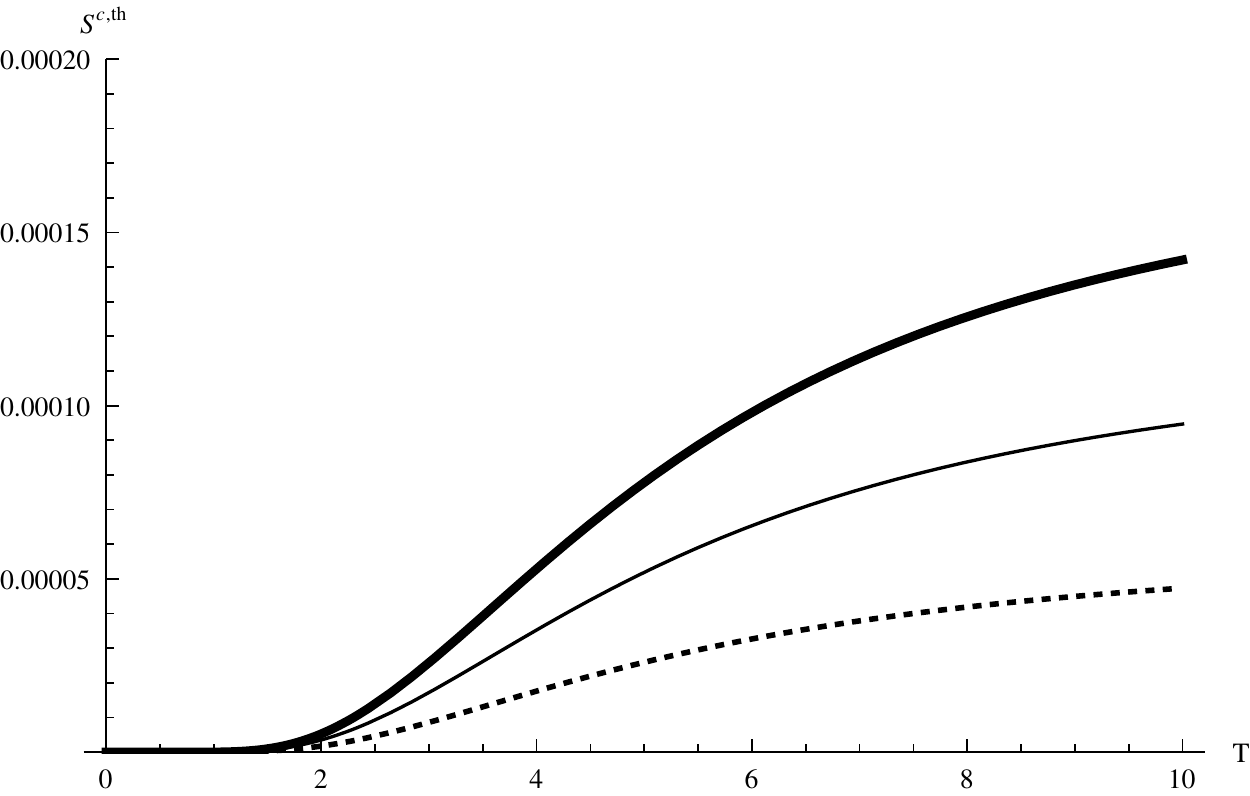}
   \caption{Plots of the entropy $S^{c,th}$ of the canonical ensemble
    vs. $T$ with $N=1$ (dotted line), $N=2$ (thin line), $N=3$ (thick line) and
    for $k_{B}=8.617\times 10^{-5}$ and $\Delta=0.001$. We see how
     additivity is clearly preserved and that the limit $T\to 0$ is
     equal to zero. }
  \label{fig:canonical}
\end{figure}

If we consider the analogous quantity for the canonical ensemble
($\hat{\rho}^{C}$ defined in Eq. (\ref{eq:36})) in the
same system, we find a very different behavior. Indeed, we can
immediately obtain that
\color{black}
\begin{equation}
  \label{eq:53}
  S^{c,th}_{N}=k_{B}\left (\frac{N \beta  \Delta  e^{-\beta  \Delta }}{e^{-\beta
      \Delta }+1}+\log \left(\left(e^{-\beta  \Delta
      }+1\right)^N\right)\right ). 
\end{equation}
\color{black}
 We can plot the first three cases for the same parameter values as
 before, and obtain an equally spaced set of functions, as we can see
 in Figure \ref{fig:canonical}.

\begin{figure}[ht]
  \centering
  \includegraphics[width=15cm]{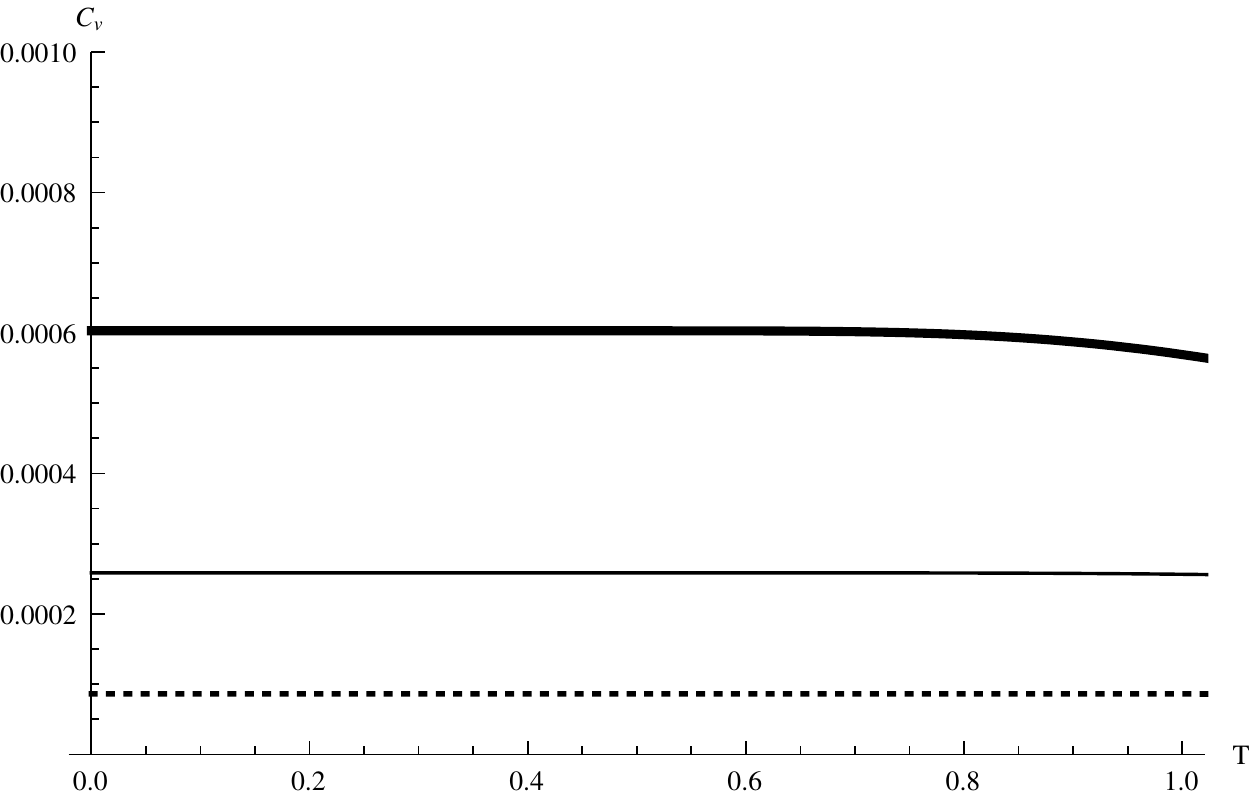}
  \caption{Plots of the specific heat of the SG ensemble vs. $T$ with
    $N=1$ (dotted line), $N=2$ (thin  line), $N=3$ (thick line)   and
     for $k_{B}=8.617\times 10^{-5}$ and $\Delta=0.001$. We can see that the limit when $T\to 0$
    increases with $N$, while for a consistent thermodynamic description it
    must go to zero.}
  \label{fig:cv}
\end{figure}

We can also consider other magnitudes such as the specific heat
$C_{v}$ (computed as $C_v=-\beta^2\frac{\partial U}{\partial \beta}$),
which  
we plot in Fig.~\ref{fig:cv}. For $C_v$, we recover the dependence obtained by 
Brody and coworkers in the simplest $N=1$ case \cite{Brody1998b}, but we can 
also see (thanks to the more general calculations in this work) how $C_v$ 
scales with the number of subsystems $N$.
Finally, notice that  $C_v$ increases with $N$ in the limit $T\to 0$, while for 
a consistent thermodynamic description it must go to zero.
This is a property that is preserved in the usual canonical
ensemble. Therefore, its violation can 
not be considered a consequence of the quantum nature of the system
but an essential property of the SG ensemble that can not be
disregarded. 
\subsection{Description in terms of density matrices}

\subsubsection{Density matrices and von Neumann entropy}

Apart from the description in terms of probability distributions, we can also 
describe the systems in the previous sections by their corresponding density
matrices, using Eq.~(\ref{eq:85}). For the simple cases $N=1$ and $N=2$ we 
obtain:

\begin{equation}
  \label{eq:17}
\hat{\rho}^{SG}_{N=1}=   \left (-\frac{\beta  \Delta  e^{\beta  \Delta
    }-e^{\beta  \Delta }+1}{\beta  \Delta  \left(1-e^{\beta  \Delta 
   }\right)}\right )\frac{|E_{0}^{1}\rangle\langle E_{0}^{1}|}{\langle
E_{0}^{1}|E_{0}^{1}\rangle} 
-\left ( \frac{-\beta  \Delta +e^{\beta  \Delta }-1}{\beta  \Delta  \left(1-e^{\beta  \Delta
   }\right)}\right )\frac{|E_{\Delta}^{1}\rangle\langle E_{\Delta}^{1}|}{\langle
E_{\Delta}^{1}|E_{\Delta}^{1}\rangle},
\end{equation}
and
\begin{align}
\label{eq:19}
\notag
  \hat{\rho}^{SG}_{N=2}&=
\left (\frac{2 \beta  \Delta +(\beta  \Delta -3) \sinh (\beta  \Delta
    )+(\beta  \Delta -2) \cosh (\beta  
   \Delta )+2}{2 \beta  \Delta  (\sinh (\beta  \Delta )-\beta  \Delta
   )} \right )\frac{| E_{0}^{2}\rangle\langle E_{0}^{2}|}{\langle
E_{0}^{2}|E_{0}^{2}\rangle}  
\\\notag
&+\left (\frac{\beta ^2 \Delta ^2-2 \cosh 
   (\beta  \Delta )+2}{\beta ^2 \Delta ^2-\beta  \Delta  \sinh (\beta
   \Delta )}\right )\frac{|E_{\Delta}^{2a}\rangle\langle E_{\Delta}^{2a}|}{\langle
E_{\Delta}^{2b}|E_{\Delta}^{2a}\rangle}+
\left (\frac{\beta ^2 \Delta ^2-2 \cosh 
   (\beta  \Delta )+2}{\beta ^2 \Delta ^2-\beta  \Delta  \sinh (\beta
   \Delta )} \right )\frac{| E_{\Delta}^{2b}\rangle\langle E_{\Delta}^{2b}|}{\langle
E_{\Delta}^{2b}|E_{\Delta}^{2b}\rangle} 
\\
&+\left (\frac{2 \beta  \Delta 
   -(\beta  \Delta +3) \sinh (\beta  \Delta )+(\beta  \Delta +2) \cosh
   (\beta  \Delta )-2}{2 \beta  \Delta  
   (\beta  \Delta -\sinh (\beta  \Delta ))}\right)\frac{|E_{2\Delta}\rangle\langle E_{2\Delta}|}{\langle
E_{2\Delta}|E_{2\Delta}\rangle}, 
\end{align}
where $|E_{0}^{1}\rangle$ and $|E_{\Delta}^{1}\rangle$ represent the energy eigenstates 
for $N=1$, and $|E_{0}^{2}\rangle, |E_{\Delta}^{2a}\rangle, |E_{\Delta}^{2b}\rangle$ 
and $| E_{2\Delta}\rangle$ are the eigenstates associated with the $N=2$ case.
The vectors $|E_{\Delta}^{2a}\rangle$ and $|E_{\Delta}^{2b}\rangle$ span the
two-dimensional eigenspace with energy equal to $E_{\Delta}$ in the
last $N=2$ case.

It is  now  easy to verify that 
\begin{equation}
  \label{eq:14}
 \hat \rho^{SG}_{N=2}\neq\hat \rho^{SG}_{N=1}\otimes \hat \rho^{SG}_{N=1}.
\end{equation}
It suffices to check that the coefficients of the projectors onto the ground 
state ($\frac{|E_{0}\rangle\langle E_{0}|}{\langle E_{0}|E_{0}\rangle}$) do not 
coincide. This relation translates Eq.~(\ref{eq:63}) into the language of 
density matrices.
Indeed, since the Hamiltonian does not introduce a coupling
among the different subsystems, the tensor product of equilibrium density 
states representing single subsystems (such as Eq.~(\ref{eq:17})), should
define equilibrium density states of the composite system. This is a property 
that holds for the usual canonical ensemble in Eq.~(\ref{eq:36}), but fails 
for the SG distribution, as we have just shown.

\begin{figure}[ht]
      \centering
\includegraphics[width=15cm]{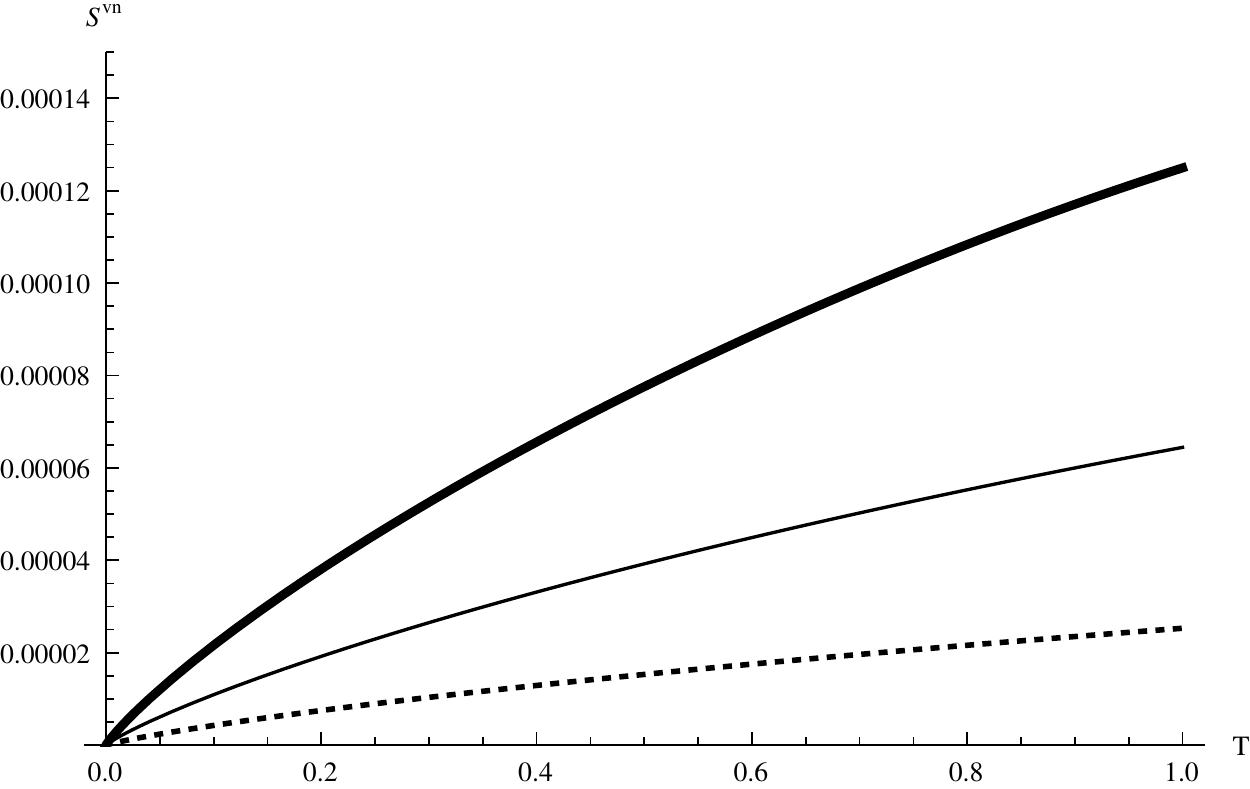}
      \caption{Plot of von Neumann's entropy of the SG ensemble for
        $N=1$ (dotted line), $N=2$
        (thin line) and $N=3$ (thick line), vs.~$T$, with  $k_{B}=8.617\times
        10^{-5}$ and $\Delta=0.001$. We can notice that the three
        levels are not equally spaced}
  \label{fig:vN3}
    \end{figure}

The SG distribution can also be studied from the point of view of von Neumann's 
entropy  
\begin{equation}
  \label{eq:18}
  S^{vn}=-k_{B}\mathrm{Tr}(\hat \rho^{SG}\log
\hat \rho^{SG}).
\end{equation}
We know that this quantity is always positive and well defined for any density
matrix we evaluate it on. We can also plot (see Fig.~\ref{fig:vN3}) the von Neumann entropies for $N=1$, 
$N=2$ and $N=3$ and verify that they are not equally spaced as 
in the canonical distribution (Figure \ref{fig:canonical}),
where the thermodynamic and the von Neumann entropy functions coincide.
Thus,  as a
consequence of the non-factorizability of the  
partition function, the von Neumann entropy is not additive, even if
the subsystems are non-interacting. 

Obviously, by comparing Figure ~\ref{fig:vN3} 
and Figure ~\ref{fig5} we can easily verify that, in the SG case, 
von Neumann entropy is not the same function as $S^{th}$ while in the
canonical distribution both entropies coincide.

\subsubsection{Entanglement}

We can explain this non-additivity of von Neumann's entropy in a simple manner 
if we appeal to the notion of entanglement. If we compute the partial trace
of the density matrix in Eq.~(\ref{eq:19}), we obtain a density matrix $\hat 
\rho_{1}= \mathrm{Tr}_{2}\hat \rho^{SG}_{N=2}$, which we can write as
\begin{align}
  \label{eq:20}
  \notag
  \hat \rho_{1}&= 
\left (\frac{2 \beta  \Delta  (\beta  \Delta -1)+(3-\beta  \Delta ) \sinh (\beta
    \Delta )-(\beta  \Delta +2) \cosh (\beta  \Delta )+2}{2 \beta  \Delta 
   (\beta  \Delta -\sinh (\beta  \Delta ))}\right )
\frac{|E_{0}\rangle\langle E_{0}|}{\langle
E_{0}|E_{0}\rangle} 
\\
&+\left (\frac{2 (\beta  \Delta  (\beta 
   \Delta +1)+1)-(\beta  \Delta +3) \sinh (\beta  \Delta )+(\beta  \Delta -2)
   \cosh (\beta  \Delta )}{2 \beta  \Delta  (\beta  \Delta -\sinh (\beta  \Delta
   ))}\right )
\frac{|E_{\Delta}\rangle\langle E_{\Delta}|}{\langle
E_{\Delta}|E_{\Delta}\rangle} 
\end{align}
Analogously, we can define (and compute)
\begin{equation}
  \label{eq:21}
  \hat \rho_{2}=\mathrm{Tr}_{1}\hat \rho^{SG}_{N=2}=\hat \rho_{1}.
\end{equation}

However we can immediately verify that
\begin{equation}
  \label{eq:22}
  \hat \rho^{SG}_{N=2}\neq \hat \rho_{1}\otimes \hat \rho_{2}.
\end{equation}

This is just reflecting upon the fact that SG density matrices, $\hat 
\rho^{SG}$, represent entangled states, as we advanced in 
Sec.~\ref{sec:orig-non-extens}. Indeed, being defined from partition
functions $Z^{SG}$ which are not factorizable, the description in
terms of density  
matrices $\hat \rho^{SG}$ must also encode this property, and it does so in the
standard way. We also know that, while separable states have additive von 
Neumann entropy (see Appendix \ref{sec:addit-von-neum} for a simple proof), the von 
Neumann entropy of entangled states is never additive (the difference between 
the entropy of the composite system with respect to the sum of the
entropies of the subsystems is what is called the \textit{quantum mutual 
information} in quantum information theory).

\section{Conclusion}
\label{sec:conclusions}

In this paper we have proved that  the  Schrödinger-Gibbs density
distribution, defined as the exponential of
(minus $\beta$ times) the expected value of the Hamiltonian operator on the
corresponding ray, cannot be  used to construct a consistent quantum
thermodynamics.  It has severe problems at the thermodynamical level
(the third law of thermodynamics is not satisfied, see for instance
Figure \ref{fig:cv}), and we have 
devoted some time to show that, in particular, the thermodynamic functions
defined by it are non-extensive.

Notice that the definition of the ensemble (Eq. (\ref{eq:46})) makes
perfect sense for finite dimensional quantum systems, where the
integral over the projective space is well defined. The
framework we used in Appendix \ref{sec:partition-function} to
perform the integrals for the partition function and to determine from
there the expression of $\hat{\rho}^{SG}$ (Eq. (\ref{eq:85})) depends also
on the finiteness of the corresponding Hilbert space. A natural
question arises: can we extend
our conclusions to general quantum systems, defined on infinite
dimensional spaces? The answer is not simple since the definition
itself becomes quite subtle in infinite dimensions, because the rigorous
mathematical definition of a functional integral over an infinite dimensional projective
space is not an easy task. However let us assume that the definition of the
functional integral is done and  let us focus on the generalization of
the results. There are two important points to discuss: the procedure
we used in  Appendix \ref{sec:partition-function} and the results we obtain. 
Obviously the procedure makes sense only if the definition of the
integral is done via a limit process over finite dimensional
approximations. In this case we could define an analog of our
integrals (Eq. (\ref{eq:32}-~(\ref{eq:76})) for each finite dimensional approximation without any
change. Regarding the results,  the limit process would be
straightforward as long as the spectrum of 
the Hamiltonian operator is purely discrete, i.e., if its essential
spectrum is empty (see \cite{Sternberg2005,Davies2007}).  In that
case,  the dimension of all the eigenspaces is finite 
and there are no accumulation points in the spectrum.
In a situation like this,  Eq. (\ref{eq:76b}) and E. (\ref{eq:85}) make sense even in
infinite dimensions.

We conclude thus that the SG ensemble does not make sense as an
equilibrium distribution because the thermodynamics associated with it
fails to satisfy very basic properties such as additivity or the Third
Law.  On the other hand, the 
use of density distributions is a natural option when studying hybrid 
quantum-classical models such as Ehrenfest models in nonadiabatic molecular 
dynamics (see \cite{Alonso2011} and \cite{Alonso2012a}).We
leave for future work the study of the consequences of the analysis in this paper for the 
equilibrium  and non-equilibrium statistics of hybrid quantum-classical 
systems (see \cite{Gallavotti2013b} for a  recent approach to nonequilibrium
and irreversibility), where the nonlinear effects on the dynamics produced by the classical 
subsystem may alter significantly the results we have presented here.

\appendix

\section{Additivity of von Neumann's estropy}
\label{sec:addit-von-neum}

We will include, for completeness, a classical proof of the additivity of von 
Neumann's entropy (see, for instance, \cite{Wehrl1978}).

Let ${\cal S}_{12}={\cal S}_1+{\cal S}_2$ be a composite system with Hilbert 
space $\mathcal{H}_{12}=\mathcal{H}_1\otimes \mathcal{H}_2$, of dimension 
$d_{12}$. We want to prove that the von Neumann entropy is additive on product
states $\hat\rho_{12}=\hat\rho_1\otimes\hat\rho_2$, i.e., 
\begin{equation}
    \label{eq:28}
  S(\hat\rho_{12})=S(\hat\rho_1)+S(\hat\rho_2),   
\end{equation}
where $\mathrm{Tr}\hat\rho_{12}=1$.

Additivity comes from the fact that the spectrum of
$\rho_{12}=\rho_1\otimes\rho_2$ consists of the products of the
eigenvalues of $\rho_1$ and $\rho_2$: 
\begin{equation}
  \label{eq:29}
\left\{
\begin{aligned}
  \hat\rho_1&=\sum_{i=1}^{d_1} \mu_i |\mu_i\rangle \langle \mu_i|
  \\
  \hat\rho_2&=\sum_{j=1}^{d_2} \nu_j |\nu_j\rangle \langle \nu_j|
\end{aligned}
\right.
\quad\Rightarrow\quad
\hat\rho_{12}=\sum_k r_k|r_k\rangle \langle r_k|=
\sum_{i=1}^{d_1}\sum_{j=1}^{d_2} \mu_i\nu_j |\mu_i\nu_j\rangle \langle
\mu_i\nu_j| ,
  \end{equation}
where we denote as $\{ |\mu_{i}\rangle\}_{i=1, \cdots, d_{1}}$ and $\{
|\nu_{j}\rangle \}_{j=1, \cdots, d_{2}}$ the eigenvectors of
$\hat\rho_{1}$ and $\hat\rho_{2}$ respectively, and as $\{ \mu_{i}\}_{i=1,
  \cdots, d_{1}}$ and $\{ \nu_{j}\}_{j=1, \cdots, d_{2}}$ the
corresponding eigenvalues.

If we compute von Neumann's entropy in the eigenbasis of $\hat\rho_{12}$ we obtain
\begin{align}
S(\hat\rho_{12})&=-k_{B}\sum_k r_k\log
r_k=-k_{B}\sum_{i=1}^{d_1}\sum_{j=1}^{d_2}\mu_i\nu_j \log(\mu_i\nu_j) \\
&=-k_{B}\sum_{i=1}^{d_1}\sum_{j=1}^{d_2}\mu_i\nu_j (\log\mu_i+\log\nu_j)
\nonumber \\
&= -k_{B}\sum_{j=1}^{d_2} \nu_j\sum_{i=1}^{d_1}\mu_i \log\mu_i-
k_{B}\sum_{i=1}^{d_1}\mu_i  \sum_{j=1}^{d_2}  \nu_j \log\nu_j\\
&=- k_{B}\sum_{i=1}^{d_1}\mu_i \log\mu_i-k_{B}\sum_{j=1}^{d_2} \nu_j \log\nu_j
\nonumber \\
&=S(\hat\rho_{1})+S(\hat\rho_{2}),
\end{align}
where the fact that $\mathrm{Tr}\hat\rho_1=\mathrm{Tr}\hat\rho_2=1$ has been
used. As the trace does not depend on the basis, the result is proved.

\section{Integrals on the Projective space}
\label{sec:integr-proj-space}

For the sake of completeness, we prove in this appendix the following result:
Consider a function $f$ defined on the sphere $S^{2n-1}\subset 
\mathbb{R}^{2n}$, which is constant along the fibers of the fibration 
\begin{equation}
  \label{eq:1}
  \tau:S^{2n-1}\to \mathbb{CP}^{n-1};
\end{equation}
i.e., which can be obtained as the pullback $\tau^{*}(f)$ of a function $f$ on
the projective space, or, from the physical point of view, which
represents a true physical quantity, as it does not depend on the
global phase of the state. Notice that Eq.~(\ref{eq:1}) corresponds to
the restriction of the canonical fibration $\mathbb{C}^{n}\to
\mathbb{CP}^{n-1}$ to the states of norm equal to one.

Then, we have that
\begin{equation}
  \label{eq:2}
  \int_{S^{2n-1}}d\mu_{S}\tau^{*}(f)=2\pi \int_{\mathbb{CP}^{n-1}}d\mu_{C}f,
\end{equation}
where $d\mu_{S}$ and $d\mu_{C}$ represent the corresponding volume forms.

Let us recall that both the sphere and the projective space are
nontrivial differentiable manifolds and therefore that integrals on
them are obtained by patching together the integrals on the charts of
their atlases. Thus, given an open covering $\{U_{k}\}$ for the
manifold $M$, we consider a  subordinated partition of unity $\{ \epsilon_{k}\}$,
i.e., a collection of functions $\epsilon_{k}:U_{k}\to \mathbb{R}$
that satisfy that for any point $p\in M$, the sum of the functions
corresponding to the open sets to which the point belong, is equal to
one:
\begin{equation}
  \label{eq:3}
  \forall p\in U_{j_{1}}\cap \cdots \cap U_{j_{m}}\Rightarrow \sum_{k=1}^{m}\epsilon_{j_{k}}(p)=1.
\end{equation}
With these, we define the integral on the manifold $M$ as
\begin{equation}
  \label{eq:4}
  \int_{M}d\mu(p) F(p)=\sum_{k}\int_{U_{k}}d\mu_{U_{k}}(p)\epsilon_{k}(p)F_{U_{k}}(p),
\end{equation}
where $F$ represents a function on $M$, $d\mu$ represents the corresponding
volume element, and $d\mu_{U_{k}}$ and $F_{U_{k}}$ represent the restrictions 
of the volume element and the function, respectively, to the open set $U_{k}$. 

We can now use the bundle structure $\tau:S^{2n-1}\to \mathbb{CP}^{n-1}$
to define a covering for $S^{2n-1}$ as a product:
\begin{equation}
  \label{eq:10}
  V^{S^{2n-1}}_{lm}=U^{S^{1}}_{l}\times U^{\mathbb{CP}^{n-1}}_{m},
\end{equation}
where $\{U^{S^{1}}_{l}\}$ represent open sets of a covering for
$S^{1}$ and $\{ U^{\mathbb{CP}^{n-1}}_{m}\}$ represent open sets of
a covering of the projective space. Such a covering always exists
because of the bundle structure. 

Then, consider a partition of unity
$\{ \epsilon_{k}\}$
for $\mathbb{CP}^{n-1}$ associated with the covering defined by the open
sets $\{ U^{\mathbb{CP}^{n-1}}_{k}\}$. We can extend this family to
define a covering for $S^{2n-1}$ by considering a partition of the
unity $\{ \varepsilon_{1}, \varepsilon_{2}\}$ associated with the
covering $\{W_{1}, W_{2}\}$ for the group $U(1)\sim
S^{1}$, and defining
\begin{equation}  \label{eq:23}
  \omega_{jk}(p)=\varepsilon_{j}(p)\epsilon_{k}(p).
\end{equation}
It is trivial to verify that $\{ \omega_{jk}\}$ defines a partition of
the unity related to the covering $\{ V^{S^{2n-1}}_{jk}\}$.

Next, we can define the integral on the sphere as:
\begin{equation}
  \label{eq:24} \int_{S^{2n-1}}d\mu_{S}\tau^{*}(f)=\sum_{jk}\int_{V^{S^{2n-1}}_{jk}}d\mu_{V_{jk}}\omega_{jk}\tau^{*}(f), 
\end{equation}
and, as the integrand is constant on the fibers, we can split the integral in
the following way:
\begin{equation}
  \label{eq:24b}
  \int_{S^{2n-1}}d\mu_{S}\tau^{*}(f)=\left (\sum_{j}\int_{j}d\mu_{W_{j}}\varepsilon_{j} \right )\sum_{k}\int_{U^{\mathbb{CP}^{n-1}_{k}}}d\mu_{U_{k}}\epsilon_{k}\tau^{*}(f). 
\end{equation}
By definition the integral on $U(1)$ is equal to $2\pi$ and thus we
proved that Eq.~(\ref{eq:2}) holds.

\section{The partition function}
\label{sec:partition-function}

Our goal in this appendix is to compute the partition function
\begin{equation}
  \label{eq:60}
  Z^{SG}=  \int_{S}d\mu(|\psi\rangle)e^{-\beta\langle \psi |\hat
    H\psi\rangle},
\end{equation}
where we recall that the relation of the integral on the unit sphere
and the integral on the projective space (which is the physically
meaningful one)   is explained in Appendix \ref{sec:integr-proj-space}.
Consider this distribution written as in Eq.~(\ref{eq:46}) and implement the 
constraint by a complex integral in the form
\begin{align}
  \label{eq:32}\notag
  Z^{SG}&=\frac 1{2\pi}\int_{-\infty}^{\infty}d\lambda
  \int_{\mathbb{C}^{n}}d\mu(\psi)e^{-\beta
    \sum_{k}E_{k}|\psi_{k}|^{2}} e^{i\lambda
    (\langle\psi|\psi\rangle-1)}
	\\
	&= \frac 1{2\pi}\int_{-\infty}^{\infty}d\lambda
  e^{-i\lambda }
  \int_{\mathbb{C}^{n}}d\mu(\psi)\prod_{k= 1}^{n}e^{-(\beta
    E_{k}-i\lambda)|\psi_{k}|^{2}  } ,
\end{align}
where $d\mu(\psi)=\prod_{k= 1}^{n}d\psi_{k} d\bar \psi_{k}$ is the canonical
volume element in $\mathbb{C}^{n}$.

Now, the Gaussian integrals factorize and can be computed straightforwardly:
\begin{align}
  \label{eq:34}
  Z^{SG}&=\frac 1{2\pi}\int_{-\infty}^{\infty}d\lambda e^{-i\lambda}
  \prod_{k= 1}^{n}\int_{\mathbb{C}}d\mu (\psi_{k})e^{-(\beta
    E_{k}-i\lambda)|\psi_{k}|^{2}}\nonumber \\
 &=\frac 1{2\pi}\int_{-\infty}^{\infty}d\lambda e^{-i\lambda}
  \prod_{k= 1}^{n} \frac{2\pi}{(\beta E_{k}-i\lambda)}.
\end{align}
If we also take into account the degeneracies of the eigenvalues we have:
\begin{equation}
  \label{eq:78}
  Z^{SG}=(2\pi)^{n-1}\int_{-\infty}^{\infty}d\lambda e^{-i\lambda}
 \color{black}\prod_{k=0}^{p}\color{black}\frac{1}{(\beta E_{k}-i\lambda)^{d_{k}}}\, ,
\end{equation}
where $d_{k}$ represents the degeneracy of eigenvalue $E_{k}$ and the
product runs only over different eigenvalues (we assume that there
are  $p+1$  of them).

The last integral (over $\lambda$) must be evaluated on the complex plane,
where the integration runs over the real axis. To this end, let us first make a
change of coordinates
\begin{equation}
  \label{eq:81}
  \xi=i\lambda,
\end{equation}
that produces:
\begin{equation}
  \label{eq:82}
  Z^{SG}=
  i(2\pi)^{n-1}\int_{\gamma}d\xi e^{-\xi} 
 \color{black}\prod_{k=0}^{p}\color{black}\frac{1}{(\beta
   E_{k}-\xi)^{d_{k}}}\, ,
\end{equation}
where now $\gamma$ is the imaginary axis.

\begin{figure}[ht]
  \centering
  \includegraphics[width=10cm]{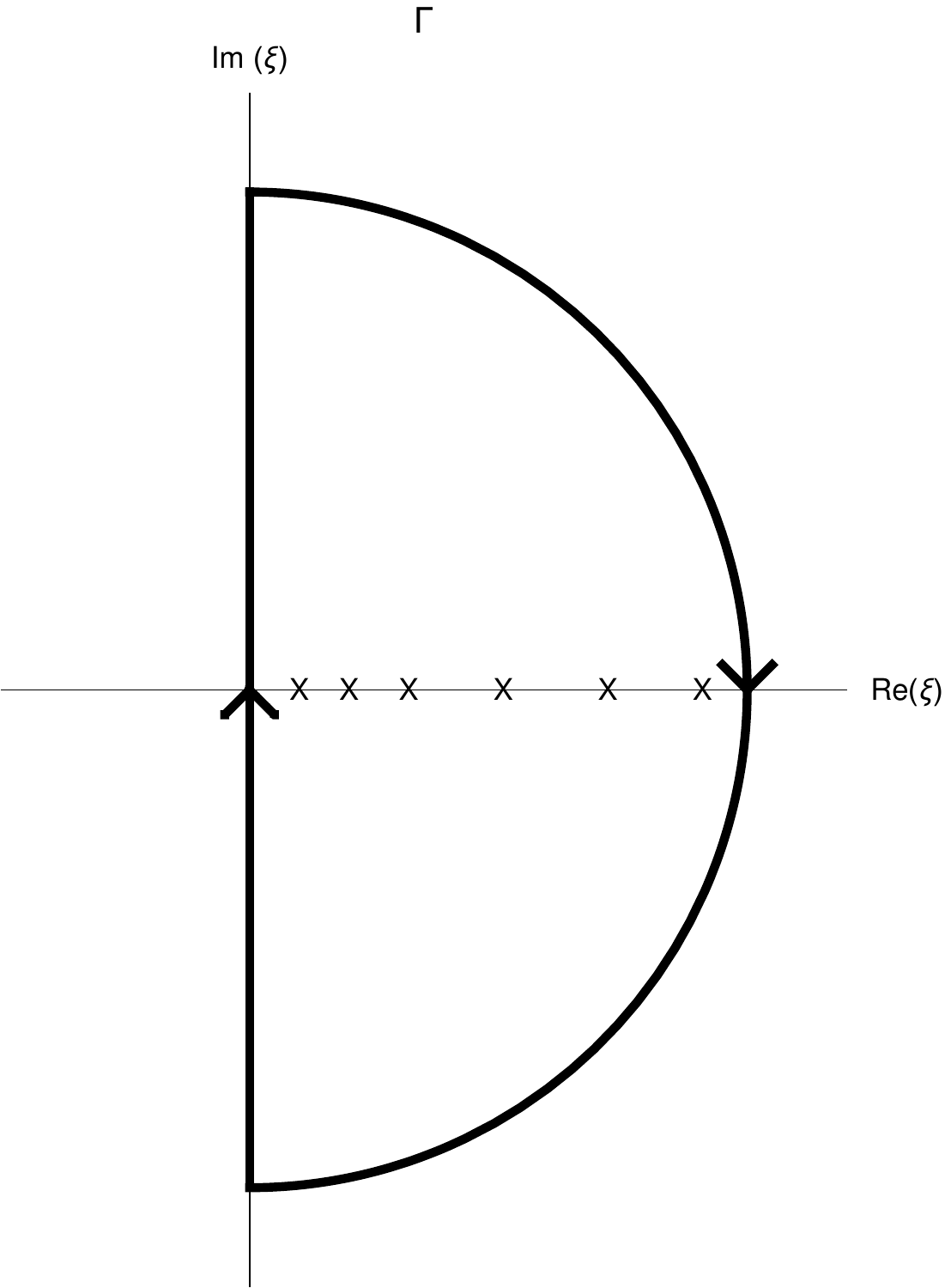}
  \caption{Representation of the integration region $\Gamma$ in the
    complex plane. The poles are represented on the real axis.}
  \label{fig:10}
\end{figure}

Next, we define a closed integration region $\Gamma$ on the left half plane 
containing the imaginary axis and a semicircle of infinite radius as it is 
depicted in Fig.~\ref{fig:10}. 
From the Cauchy theorem, we know that the integral is equal to the sum of the 
corresponding residues at the poles of the integrand, which in this case lie on
\begin{equation}
  \label{eq:73}
  \xi_{k}=\beta E_{k},
\end{equation}
their order corresponding to the degeneracy of the corresponding eigenvalue.
We have thus that:
\begin{equation}
  \label{eq:83}
  Z^{SG}=i(2\pi)^{n-1} 2\pi i\color{black}\sum_{j=0}^{p}\color{black}\mathrm{Res}\left
    (\frac{e^{-\xi}}{\prod_{k=1}^{p}(\beta E_{k}-\xi)^{d_{k}}},\xi=\beta E_{j} \right ) .
\end{equation}

Then, we can write:
\begin{equation}
  \label{eq:76}
  Z^{SG}=\color{black}\sum_{k=0}^{p}\color{black}e^{-\beta E_{k}}F_{E_{k}},
\end{equation}
where 
\begin{equation}
  \label{eq:77}
  F_{E_k}=-(2\pi)^{n} e^{\beta E_{k}} \frac 1{(d_{k}-1)!}\left.\left
  [\frac{\partial^{d_{k}-1} }{\partial \xi^{d_{k}-1} 
  }\left (e^{-\xi}\prod_{\substack{s=
  0\\ s\neq k}}^{p} \frac{1}{(\beta E_{s}-\xi
  )^{d_{s}}}\right )\right] \right|_{\xi= \beta E_{k}}.
\end{equation}

By computing the derivative we obtain
\color{black}
\begin{equation}
  \label{eq:80}
  F_{E_k}=-\frac {(2\pi)^{n}}{(d_{k}-1)!}\sum_{\substack{j_{0}, \cdots,
    j_{p}= 0\\ j_{0}+j_{1}+\cdots
    j_{p}=d_{k}-1}}^{d_{k}-1}
\binom{d_{k}-1}{j_{0},\cdots, j_{p}}
\prod_{\substack{s=
  0\\ s\neq k}}^{p}\frac{(-1)^{j_{0
  }}\frac{(d_{s}+j_{s}-1)!}{(d_{s}-1)!}}{(\beta E_{s}-\beta
E_{k})^{d_{s}+j_{s}}}\, ,
\end{equation}
\color{black}
where $\binom{d_{k}-1}{j_{0}, \cdots, j_{s}}$ represents the
multinomial coefficient
$$
\binom{d_{k}-1}{j_{0}, \cdots, j_{s}}=\frac{(d_{k}-1)!}{j_{0}! \cdots
  j_{s}!}\, .
$$
This concludes the proof.


\begin{acknowledgments}
 We would like to thank Professor Mario Floría for several
enlightening discussions that guided us in the right direction.
We would also like to thank an anonymous referee for a
very careful reading of the manuscript, which has improved
it in a very significant manner. This work was supported by
Diputación General de Aragón (SPAIN) Grants No. E24/1, No. E24/2, and No. E24/3;
Banco de Santander and Universidad de Zaragoza  Grant No. UZ2012-CIE-06; and
Ministerio de Economia y Competitividad (Spain)  Grants No. FIS2013-46159-C3-2-P and No.
FPA2012-35453.

\end{acknowledgments}


%

\end{document}